\documentclass[12pt]{article}
\usepackage{graphicx} 
\begin{document}
\centerline{\bf Three-state majority-vote model
on square lattice}

\bigskip
\centerline{ F. W. S. Lima}
\bigskip

Dietrich Stauffer Computational Physics Lab, Departamento de F\'{\i}sica, \\
Universidade Federal do Piau\'{\i}, 64049-550, Teresina - PI, Brazil.\\
\medskip

e-mail: fwslima@gmail.com

\bigskip

{\small Abstract:Here, the model of non-equilibrium model with two states ($-1,+1$) and a noise $q$ on simple square lattices proposed for M.J. Oliveira (1992) following the conjecture of up-down symmetry of Grinstein and colleagues  (1985) is studied and generalized. This model is well-known, today, as Majority-Vote Model. They showed, through Monte Carlo simulations, that their obtained results  fall into the universality
class of the equilibrium Ising model on a square lattice. In this work, we generalize the Majority-Vote Model for a version with three states, now including the zero state, ($-1,0,+1$) in two dimensions. Using Monte Carlo simulations, we showed that our model falls into the universality class of the spin-$1$ ($-1,0,+1$) and spin-$1/2$
Ising model  and also agree with Majority-Vote Model proposed for M.J. Oliveira (1992) . The exponents ratio obtained for our model was $\gamma/\nu =1.77(3)$, $\beta/\nu=0.121(5)$, and $1/\nu =1.03(5)$. The critical noise obtained and the fourth-order cumulant were $q_{c}=0.106(5)$ and $U^{*}=0.62(3)$.}
\bigskip

\section{Introduction}
The Ising model \cite{a3,onsager} has been used during long time as a "toy model" for diverses objectives, as to test and to improve new algorithms and methods of high precision for calculation of critical exponents in Equilibrium Statistical Mechanics  using the Monte Carlo method
as Metropolis \cite{me}, Swendsen-Wang \cite{s-w}, Wang-Landau \cite{w-l} algorithms, Single histogram  \cite{f-s} and Broad histogram \cite{pmco} methods. The Ising model was already applied decades ago to explain how city populations segregate \cite{Schel}, how a school of fish aligns into one direction for swimming \cite{callen} or how workers decide whether or not to go on strike \cite{galam,galam1}. In the Latan\'e model of Social Impact \cite{latané} the Ising model has been used to give a consensus, a fragmentation into many different opinions, or a leadership effect when a few people change the opinion of lots of others. To some extent the voter model of Liggett \cite{ligg} is an Ising-type model: opinions follow the majority of the neighbourhood. All these cited model and others can be found out in \cite{book}. 

Realistic economics models of tax evasion appear to be necessary, because tax evasion remains to be a major predicament facing governments \cite{K,JA,L,JS}. Experimental evidence provided by G\"achter \cite{Ga} indeed suggests that tax payers tend to condition their decision regarding whether to pay taxes or not on the tax evasion decision of the members of their group. Frey and Torgler \cite{FT} also provide empirical evidence on the relevance of conditional cooperation for tax morale. Following the same context, recently, Zaklan et al. \cite{zaklan,zaklan1} developed an economics model to study the problem of tax evasion dynamics using the Ising model through Monte-Carlo simulations with the Glauber and heatbath algorithms (that obey detailed-balance equilibrium) to study the proposed model.
G. Grinstein et al. \cite{g} have argued that nonequilibrium stochastic spin systems on regular square lattices (SL) with up-down symmetry fall into the universality
class of the equilibrium Ising model \cite{g}. This conjecture was confirmed for various Archimedean lattices and 
in several models that do not obey detailed balance \cite{C,J,M,mario,lima01,lima02}. The majority-vote model with two states (MV2) is a nonequilibrium model proposed by M.J. Oliveira in $1992$ and defined by stochastic dynamics with local rules and with up-down symmetry on a regular lattice shows a second-order phase transition with critical exponents $\beta$, $\gamma$, $\nu$ which characterize  the system in the vicinity of the phase transition identical \cite{mario,a1} with those of the equilibrium Ising model \cite{a3} for regular lattices. Lima et al. \cite{lima0} 
studied MV2  on Voronoi-Delaunay random lattices
with periodic boundary conditions. These lattices posses natural quenched
disorder in their connections. They showed that presence of quenched 
connectivity disorder is enough to alter the exponents $\beta/\nu$
and $\gamma/\nu$ from the pure model and therefore that quenched disorder is a relevant term to
such non-equilibrium phase-transition which disagree with the arguments of G. Grinstein et al. \cite{g}.

Recently, simulations on both {\it undirected} and {\it directed} scale-free 
networks \cite{newman,sanchez,ba1,alex,sumour,sumourss,lima}, random graphs \cite{erdo,er2} and social networks \cite{er,er1,DS}, have attracted interest of researchers from various areas. These complex networks have been 
studied extensively by Lima et al. in the context of magnetism (MV2, Ising, and Potts model) \cite{lima1,lima2,lima3,lima4,lima5,lima6}, econophysics models \cite{zaklan1,lima8} and sociophysics model \cite{lima9}. 
Lima \cite{lima03} make an analysis of tax evasion dynamics with the Zaklan model on two-dimensional SL using MV2 for their temporal evolution under different enforcement regimes. They showed that the MV2 model also is capable to control the different levels of the tax evasion as it was made by Zaklan et al. \cite{zaklan1} using Ising model on various structures: SL, Voronoi-Delaunay random lattice, Barab\'asi-Albert (AB) network  and Erd\"os-R\'enyi (ER) graphs; we discuss the resulting tax evasion dynamics. Perhaps, this is the first application of MV2 model to a real system, in this case applied to the economy. The generalization to a three-state majority-vote model(MV3) on a regular SL was considered by \cite{tome1,tome2}, where the authors found $q_{c}=0.117(1)$. The resulting critical exponents for this non-equilibrium MV3 model are in agreement with the ones for the equilibrium three-states Potts model \cite{potts}, supporting the conjecture G. Grinstein et al. \cite{g}.
Recently, Melo et al. \cite{melo} also studied the MV3 model\cite{tome1,tome2}, now, on ER graphs \cite{erdo}. In MV3 model on ER graphs they found that the critical noise $q_{c}$ is a function of the mean connectivity $z$ of the graph. The critical exponents ratio $\gamma/\nu$, $\beta/\nu$ and $1/\nu$ was calculated for various values of connectivity. In this, Melo et al.\cite{melo} suggest a future work on the two- and three-state Potts model on random graphs would be of interest in order to provide a direct comparison with our results in light of the conjecture by Grinstein et al.\cite{g}, which states that reversible and irreversible models with same symmetry belong to the same universality class.

In the present work, we propose a non-equilibrium model with three states ($-1,0,+1$), called the three-state majority-vote model (MV3) in two dimensions as a generalization of MV2. Our main goal is to check the hypothesis of Grinstein  et al. \cite{g} described above. 
The remainder of our paper is organised as follows. In section 2, we present our  MV3 model and detail of the Monte Carlo simulation and calculations used in the evolution of physical quantities of MV3. In section 3 we make an analysis of simulations performed in the previous section and discuss the results obtained. And finally in section 4, we present our conclusions from the results obtained of MV3 model and present several perspectives work to be done at present.
\bigskip
\newpage
\section{Model and simulation }
On a SL where each site of the lattice is inhabited, at a time step, we consider the MV3 defined by a set of ``voters'' or spin variables $\sigma$ taking the values $\pm 1$ and $0$, situated on every node of the SL with $N=L^2$ sites.
The evolution is governed by single spin-flip like dynamics with a probability $w_i$ of $i$-th spin flip given by
\begin{equation}
w_{i}(\sigma)=\frac{1}{2}\biggl[ 1-(1-2q)\sigma_{i}S\biggl(\sum_{\delta
=1}^{k_{i}}\sigma_{i+\delta}\biggl)\biggl],
\end{equation}
where $S(x)$ is the sign $\pm 1$ of $x$ if $x\neq0$, $S(x)=0$ if $x=0$ and sum runs over the number $z=4$ of nearest neighbors of $i$-th spin. 
The control parameter $0\le q\le 1$ plays the role of the temperature in equilibrium systems and measures
the probability of parallel aligning to the majority of neighbors.
It means, that given spin $i$ adopts the majority sign of its neighbors with probability $q$ and the minority sign with probability $(1-q)$ .

To study the critical behavior of the model we define the variable $m\equiv m_{1}-m_{3}$ if $m1$, $m2$, and $m3$ are the numbers of sites with $\sigma_i = +1$, $0$ and $-1$, respectively, all normalized by $N$. In particular, we are interested in the
magnetization $M$, susceptibility $\chi$ and the reduced fourth-order cumulant $U$
\begin{equation}
M(q)\equiv \langle|m|\rangle,
\end{equation}
\begin{equation}
\chi(q)\equiv N\left(\langle m^2\rangle-\langle m \rangle^2\right),
\end{equation}
\begin{equation}
U(q)\equiv 1-\langle m^{4}\rangle/\left( 3\langle m^2 \rangle^2 \right),
\end{equation}
where $\langle\cdots\rangle$ stands for a thermodynamics average.
The results are averaged over the $N_{run}$ independent simulations. 
These quantities are functions of the noise parameter $q$ and obey the finite-size
scaling relations
\begin{equation}
M=L^{-\beta/\nu}f_m(x),
\end{equation}
\begin{equation}
\chi=L^{\gamma/\nu}f_\chi(x),
\end{equation}
\begin{equation}
\frac{dU}{dq}=L^{1/\nu}f_U(x),
\end{equation}
where $\nu$, $\beta$, and $\gamma$ are the usual critical 
exponents, $f_{i}(x)$ are the finite size scaling functions with
\begin{equation}
x=(q-q_c)L^{1/\nu}
\end{equation}
being the scaling variable.
Therefore, from the size dependence of $M$ and $\chi$
we obtained the exponents $\beta/\nu$ and $\gamma/\nu$, respectively.
The maximum value of susceptibility also scales as $L^{\gamma/\nu}$. Moreover, the
value of $q^*$ for which $\chi$ has a maximum is expected to scale with the system size as
\begin{equation}
q^*=q_c+bL^{-1/\nu}
\end{equation}
where $b\approx 1$. Therefore, the  relations  may be used to get the exponent $1/\nu$.
We performed Monte Carlo simulation on the SL with various systems of size $L=8$, $16$, $32$, $64$, $128$ and $256$.
It takes $1\times 10^5$ Monte Carlo steps (MCS) to make the system reach the steady state, and then the time averages are estimated over the next $2\times 10^5$ MCS.
One MCS is accomplished after all the $N$ spins are investigated whether they flip or not.
We carried out $N_{run}=1000$ to $10000$ independent simulation runs for each lattice and for a given set of parameters $(q,N)$.
\section{ Results and discussion} 
In Fig.1, we show the dependence of the magnetization $M$ on the noise parameter $q$, obtained from simulations on SL with lattice size $L=8,16,32,64,128$ and $256$ with $(L \times L=N)$ sites.
The shape of $M(q)$ curve, for a given value of $N$, suggests the presence of a second-order phase transition in the system. The phase transition occurs at the value of the critical noise parameter $q_c$. 
\bigskip
\begin{figure}[!hbt]
%\begin{center}
\includegraphics[angle=0,scale=0.5]{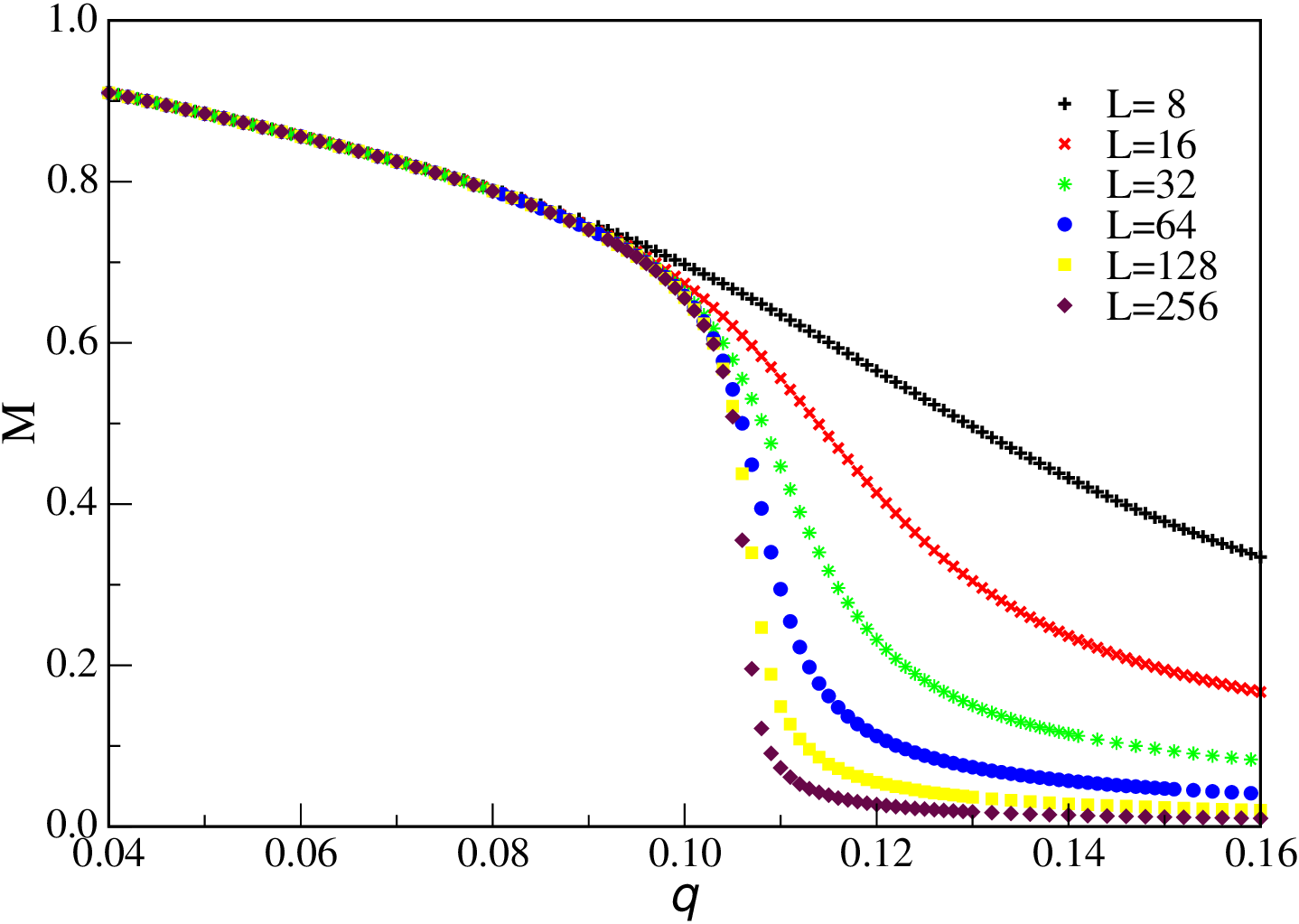}
%\end{center}
\caption{The magnetization $M$ versus the noise parameter $q$, for $L=8$, $16$, $32$, $64$, $128$ and $256$ size lattice for SL.}
\bigskip
\newpage
\end{figure} 
\begin{figure}[!hbt]
%\begin{center}
\includegraphics[angle=0,scale=0.5]{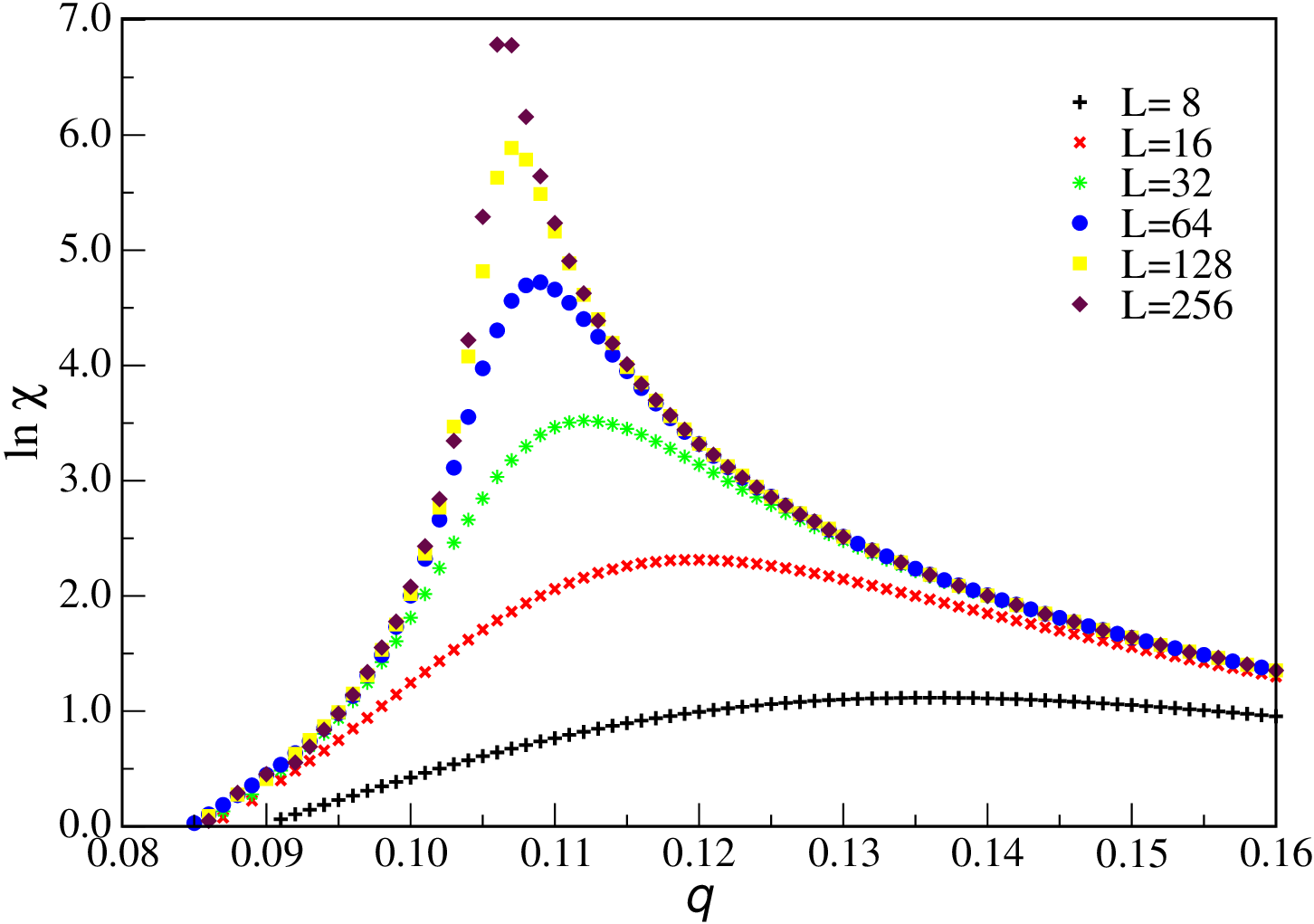}
%\end{center}
\caption {Susceptibility versus $q$ for SL. The same simulation design as in figure 1.}
\end{figure} 
In Fig. 2 the corresponding behavior of the susceptibility $\chi$ is presented.
\begin{figure}[!hbt]
%\begin{center}
\includegraphics[angle=0,scale=0.5]{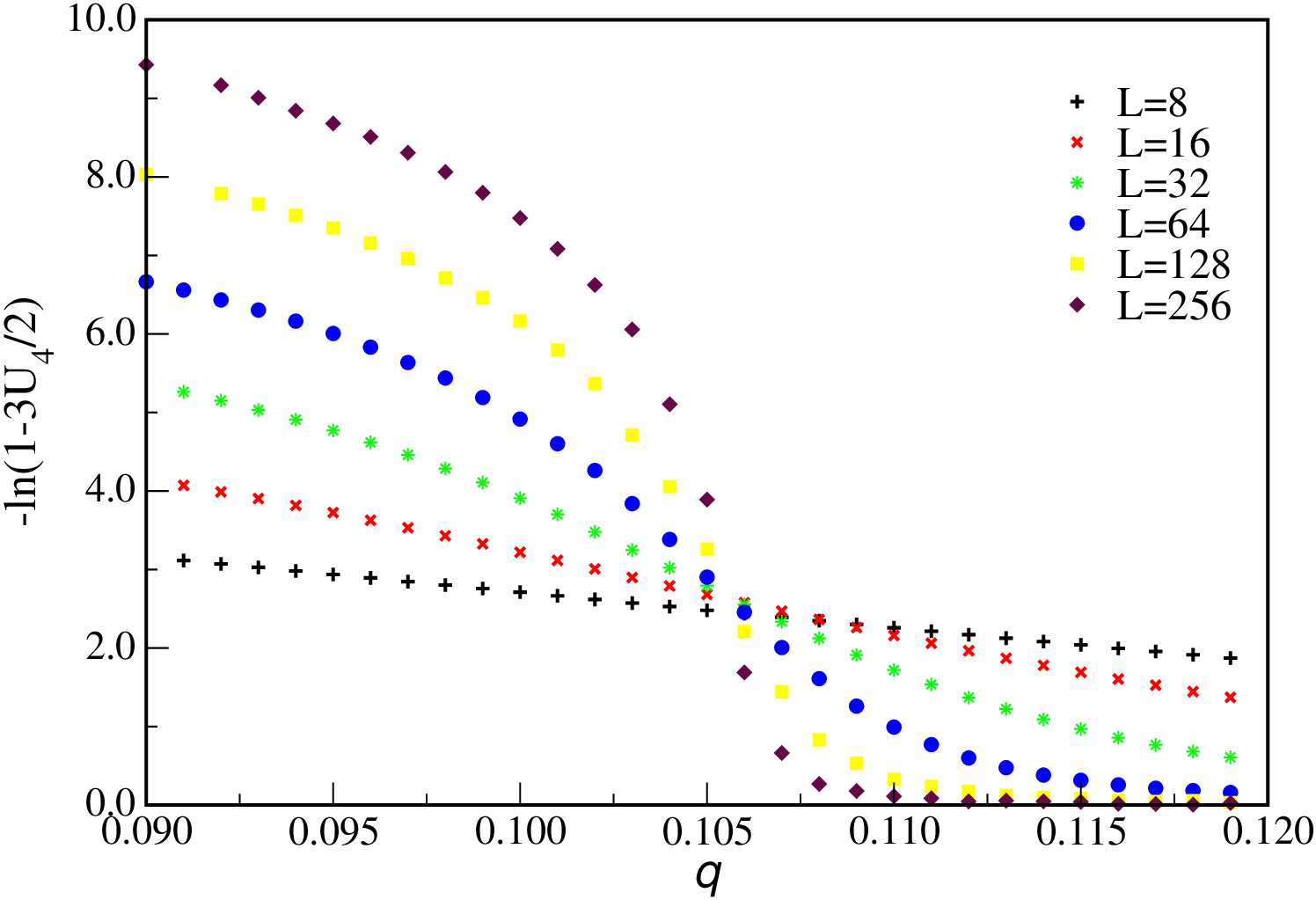}
%\end{center}
\caption{The reduced Binder's fourth-order cumulant $U$ versus $q$ for SL. The same simulation design as in figure 1.}
\end{figure}
In Fig. 3 we plot the Binder's fourth-order cumulant $U$ for different values of the system size $N$.
The critical noise parameter $q_c$ is estimated as the point where the curves for different system sizes $N$ intercept each other \cite{binder}. The critical noise obtained was $q_{c}=0.106(5)$ and "universal charge" $U^{*}=0.62(3)$.
\begin{figure}[!hbt]
%\begin{center}
\includegraphics[angle=0,scale=0.5]{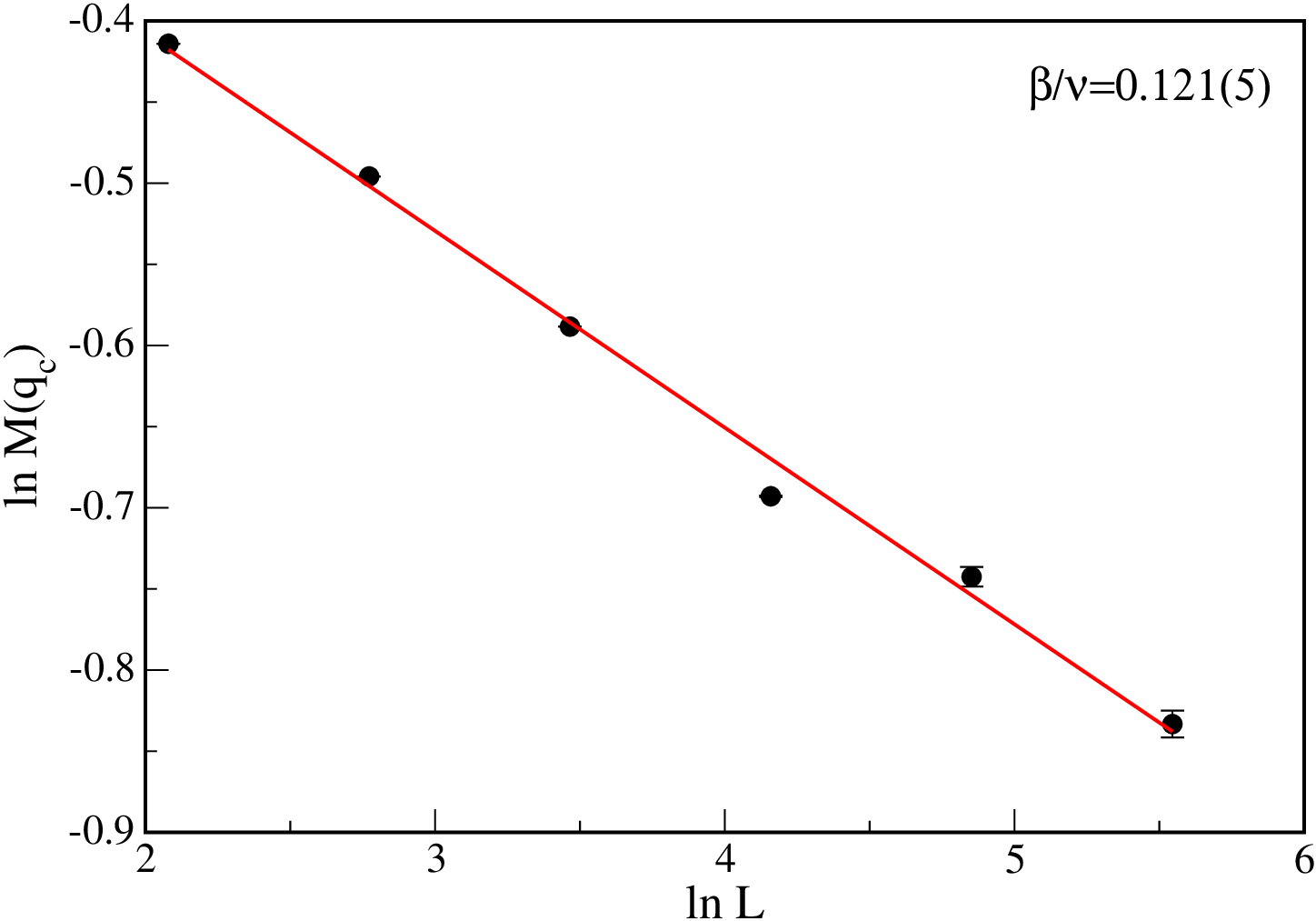}
%\end{center}
\caption{Plot of $\ln M^* $ vs. $\ln L$ for ( $L=8$, $16$, $32$, $64$, $128$ and $256$ ) SL size lattice.}
\end{figure} 
In Fig. 4 we plot the dependence of the magnetization $M^*=M(q_{c})$ versus the linear system size $L$.
The slopes of curves correspond to the exponent ratio $\beta/\nu$ according to relation (5).
The obtained exponent is $\beta/\nu=0.121(5)$.
The exponents ratio $\gamma/\nu$ are obtained from the slopes of the straight lines with $\gamma/\nu=1.77(3)$ for SL, as presented in Fig. 5 and  obtained from the relation (6).
\begin{figure}[!hbt]
%\begin{center}
\includegraphics[angle=0,scale=0.5]{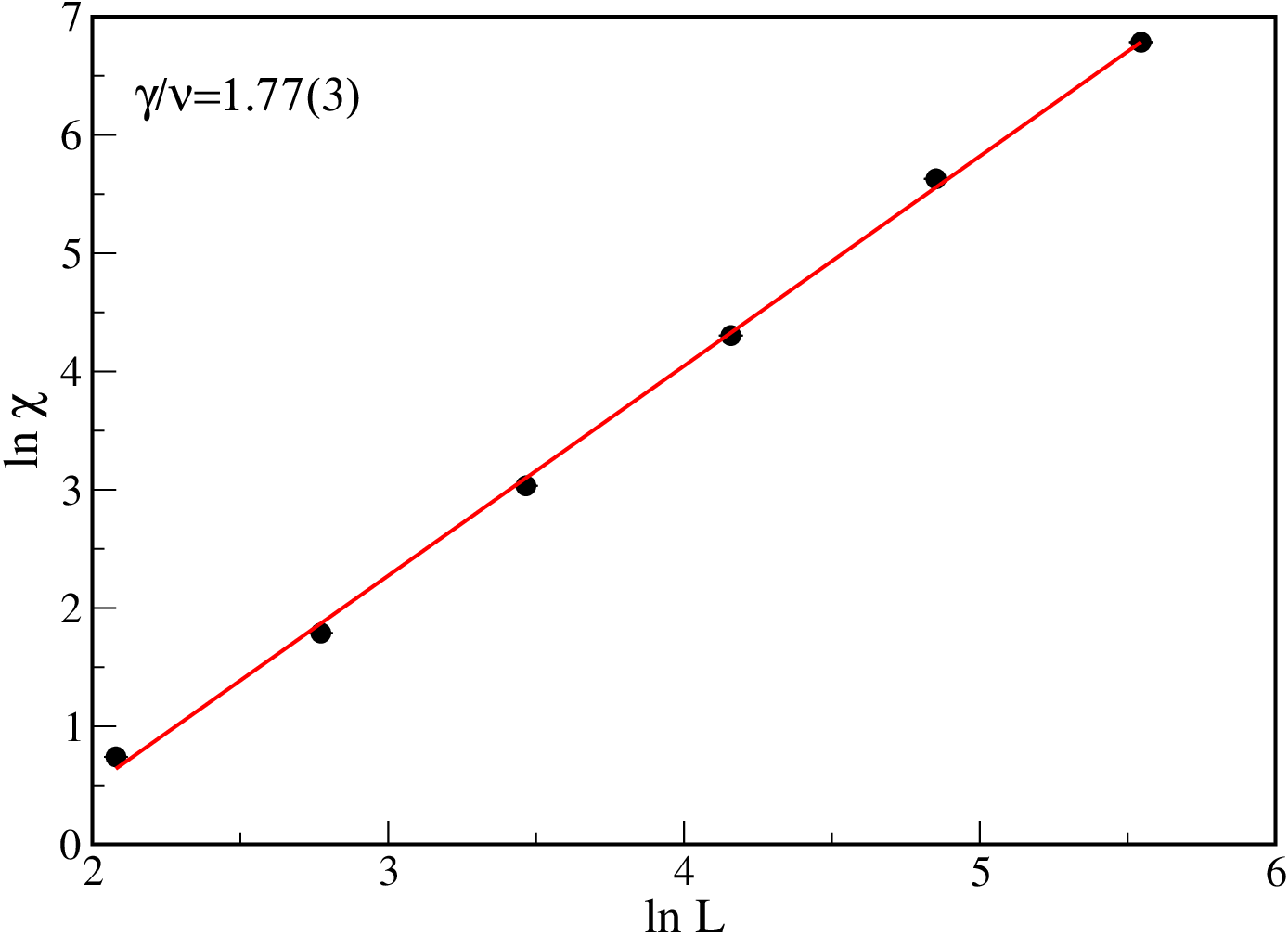}
%\end{center}
\caption{Critical behavior of the susceptibility $\chi(N)$ at $q=q_{c}$ for SL.}
\end{figure} 
To obtain the critical exponent $1/\nu$, we used the scaling relation (9).
The calculated  values of the exponents $1/\nu=1.03(5)$ . See Fig. 6.
\begin{figure}[!hbt]
%\begin{center}
\includegraphics[angle=0,scale=0.5]{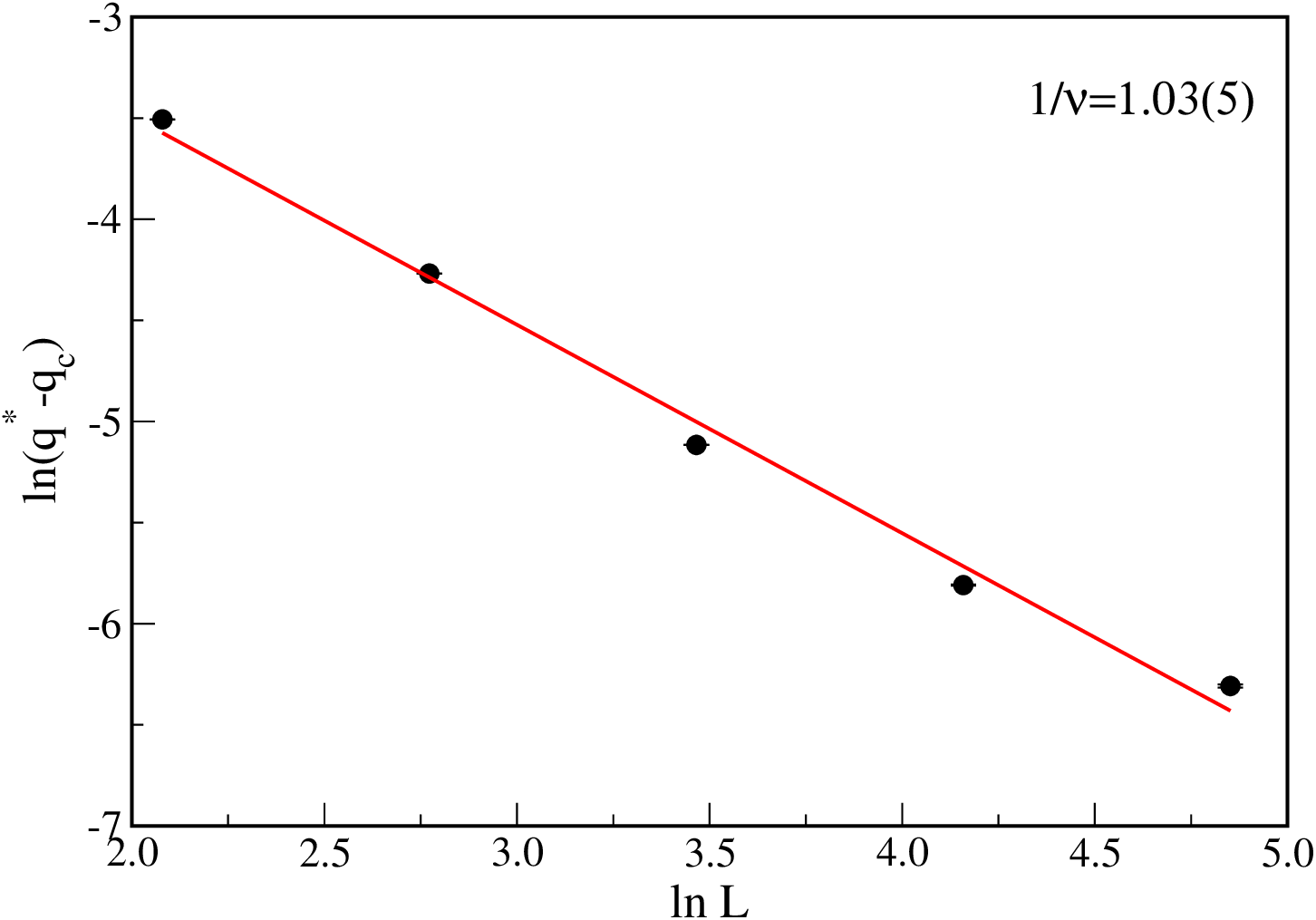}
%\end{center}
\caption{The exponents $1/\nu$ obtained from the relation (9) for SL.}
\end{figure} 

The results of simulations together with data for ER  graphs and  AB  networks with mean connectivity $\bar z=4$ are collected in Tab. 1.
\begin{table}[h]
\begin{center}
\begin{tabular}{|c||c|c|c|c|c|}
\hline
$ $ & $S$ & $\nu$ & $\alpha/\nu$& $\beta/\nu$ & $\gamma/\nu$ \\
\hline
$ (i)$ & $1/2$ & $1$ & $0$& $0.125$ & $1.75$  \\
\hline
$ (ii)$ & $1$ & $1$ & $0$ & $0.125$ & $1.75$  \\
\hline 
$(iii)$ & $1/2$ & $0.99(5),$ & $-$ & $0.125(5)$ & $1.73(5)$  \\
\hline
$(iv)$ & $1/2$ & $1.03(3)$ & $-$ & $0.123(5)$ & $1.78(5)$  \\
\hline
$(v)$ & $1$ & $0.97(5)$ & $-$ & $0.121(5)$ & $1.77(3)$  \\
\hline
\end{tabular}
\end{center}
\caption{ ({\it i}) Analytical results for ferromagnetic Ising model $2D$ with $S=1/2$ \cite{a3}, ({\it ii}) Analytical results for ferromagnetic Ising model $2D$ with $S=1$ \cite{KUTLU}. ({\it iii}) Results of M. J. Oliveira \cite{mario}. ({\it iv}) Results of Monte carlo simulations of  Kwak et al. \cite{Kwak}. ({\it v}) Our results for MV3 on SL.
} 
\end{table}
%% ----------------------------------------------------------------------------
\bigskip
\newpage
\section{Conclusion}
We presented a very simple non-equilibrium MV3 on SL.
On this lattice, the MV3 shows a second-order phase transition.
Our Monte Carlo simulations demonstrate that our MV3 model agree with the universality hypothesis of Grinstein  et al. \cite{g}.

Finally, we remark that the critical exponents $\gamma/\nu$, $\beta/\nu$ and $1/\nu$ for MV3 on SL presented in the Tab. 1 are similar the exponents from the $S=1/2$ and $S=1$  Ising model 2D \cite{a3,KUTLU}, obtained of  analytical results,   and also with the results obtained, through Monte Carlo simulation,  by  M. J. Oliveira \cite{mario} and Kwak et al. \cite{Kwak}. Therefore, our results agree with the conjecture by Grinstein et al.\cite{g}, which states that reversible and irreversible models with same symmetry belong to the same universality class. Future works using MV3 model are already in progress on {\it undirected} and {\it directed} Small-world networks and on ER random graphs.
\subsection{Acknowledgments}
The author F. W. S. Lima thank M.J. Oliveira and Dietrich Stauffer for many suggestions and fruitful discussions in the
development this work. We also acknowledge the
Brazilian agency FAPEPI (Teresina-Piau\'{\i}-Brasil) and CNPQ for  its financial support and. This
work also was supported the system SGI Altix 1350 the computational park
CENAPAD.UNICAMP-USP, SP-BRAZIL and Dietrich Stauffer Computational Physics Lab-TERESINA-PIAU\'I-BRAZIL.

\end{document}